\documentclass[epj]{svjour}
\usepackage{graphicx}
\usepackage{amsmath,amssymb}
\begin{document}
\title{Possible effects of Hybrid Gravity on stellar kinematics in elliptical galaxies}
\author{Vesna Borka Jovanovi\'{c}$^1$\thanks{\emph{e-mail:} vborka@vinca.rs}, Du\v{s}ko Borka$^1$, Predrag Jovanovi\'{c}$^2$ \and Salvatore Capozziello$^{3,4,5}$}
\authorrunning{V. Borka Jovanovi\'{c} et al.}
\institute{$^1$Department of Theoretical Physics and Condensed Matter Physics (020), Vin\v{c}a Institute of Nuclear Sciences - National Institute of the Republic of Serbia, University of Belgrade, P.O. Box 522, 11001 Belgrade, Serbia \\
$^2$Astronomical Observatory, Volgina 7, P.O. Box 74, 11060 Belgrade, Serbia \\
$^3$ Dipartimento di Fisica ''E. Pancini'', Universit\`{a} di Napoli ''Federico II'', Compl. Univ. di Monte S. Angelo, Edificio 6, Via Cinthia, I-80126, Napoli, Italy \\
$^4$ Istituto Nazionale di Fisica Nucleare (INFN) Sez. di Napoli, Compl. Univ. di Monte S. Angelo, Edificio 6, Via Cinthia, I-80126, Napoli, Italy \\
$^5$ Scuola Superiore Meridionale, Largo S. Marcellino 10, I-80138, Napoli, Italy.}
\date{Received: date / Revised version: date}
%

\abstract{We use the Fundamental Plane of Elliptical Galaxies to constrain the so-called Hybrid Gravity, a modified theory of gravity where General Relativity is improved by further degrees of freedom of metric-affine Palatini formalism of $f(\cal R)$ gravity. Because the Fundamental Plane is connected to the global properties of elliptical galaxies, it is possible to obtain observational constraints on the parameters of Hybrid Gravity in the weak field limit. We analyze also the velocity distribution of elliptical galaxies comparing our theoretical results in the case of Hybrid Gravity with astronomical data for elliptical galaxies. In this way, we are able to constrain the Hybrid Gravity parameters $m_\phi$ and $\phi_0$. We show that the Fundamental Plane, i.e. $v_c/\sigma$ relations, can be used as a standard tool to probe different theories of gravity in the weak field limit. We conclude that Hybrid Gravity is able to explain elliptical galaxies with different stellar kinematics without the dark matter hypothesis. 
} 

\PACS{
      {}{Modified  gravity} \and {}{Elliptical galaxies} \and {}{Fundamental plane.}
     } 
\maketitle

\section{Introduction}
\label{sec01}

The Fundamental Plane (FP) is connected to the global properties of elliptical galaxies. It is represented by: the central projected velocity dispersion $\sigma_0$, the effective radius $r_e$, and the mean effective surface brightness (within $r_e$) $I_e$ \cite{dres87,ciot96}. Elliptical galaxies are usually confined in a narrow logarithmic plane of their configuration space which is referred to as the FP \cite{dres87,ciot96}. It is possible to show that any of the three parameters may be estimated from the other two, and together they describe a plane that falls within a more general three-dimensional phase-space. In other words, the FP is considered a bi-variate manifold in a parametric space. The FP is defined and discussed in detail in several papers, see e.g \cite{bend92,bend93,busa97,saul13,tara15,terl81,capp13,dono13,binn98} and references therein. This important empirical relation is given by the following equation \cite{busa97}:
\begin{equation}
\log(r_e) = a \, \log(\sigma_0) + b \, \log(I_e) + c,
\label{equ11}
\end{equation}
\noindent with $a$ and $b$ being the FP coefficients which, in general, are fixed by observations. This relation gives us the possibility to obtain observational constraints on the structure, formation, and evolution of early-type galaxies. Reversing the argument, the FP can be adopted to fix parameters of a given theory of gravity, once they are constrained by observations.

In this paper, we shall adopt Eq.\eqref{equ11} to constrain parameters of the so-called Hybrid Gravity, which is a recently proposed theory of gravity \cite{capo13a} where further degrees of freedom coming from extensions of General Relativity \cite{capo11} are represented in the metric-affine formalism to cure shortcomings of $f(R)$ gravity both in metric and Palatini formalism.  

Specifically, in order to describe the velocity of stellar populations, one can define rotational velocity of a group of stars $v_c$. Also, one can define a dispersion $\sigma$ which represents the characteristic random velocity of stars. The obtained ratio $v_c/ \sigma$ is a relation which characterizes the kinematics of galaxies. In case of spiral galaxies (kinematically cold systems) $v_c/ \sigma \gg 1$, while elliptical galaxies (kinematically hot systems) are characterized by $0 < v_c/ \sigma < 1$. It is the main characteristic which differentiates spiral from elliptical galaxies.

Here, we study Extended Theories of Gravity using astronomical observations for FP. In this way, we want to give constrains to parameters of Hybrid Gravity. Extended Gravity is presented in several review  papers like 
\cite{capo11,noji11,noji17,capo13c,capo19}. Some experimental limits related to  Extended Theories of Gravity are reported in  \cite{avil12,duns16,tino20,capo15} and references therein.

The content of this paper is as follows. In Section 2, we present basics of Hybrid Gravity. In Section 3, we recover the FP of elliptical galaxies in the framework of this theory. In Section 4, we find constraints on Hybrid Gravity parameters by FP and we study the $v_c/ \sigma$ relation in the $(m_\phi,\phi_0)$ parameter space. Section 5 is devoted to conclusions.

\section{A summary of Hybrid Gravity}
\label{sec02}

Let us present here the basic formalism for  Hybrid Metric-Palatini Gravity within the equivalent scalar-tensor representation (we refer the reader to \cite{capo13a,capo13b,koiv10,capo12,bork16a} for more details). The action for Hybrid Gravity is given by
\begin{equation}
S = \int d^4 x \sqrt{-g} \left[ R + f(\mathcal{R}) + 2\kappa^2 \mathcal{L}_m \right]\, ,
\label{equ21}
\end{equation}
where $\kappa^2\equiv 8\pi G$, $R$ is the Einstein-Hilbert term defined with the Levi-Civita connection, ${\cal R } \equiv g^{\mu\nu}{\cal R}_{\mu\nu} $ is the Palatini curvature with the connection $\hat{\Gamma}^\alpha_{\mu\nu}$ independent of the metric $g_{\mu\nu}$. As discussed in \cite{capo13a}, it is possible to recast such an action as

\begin{equation}
S = \frac{1}{2\kappa^2}\int d^4 x \sqrt{-g} \left[R + \phi{\cal R}-V(\phi)\right] +S_m \,,
\label{equ22}
\end{equation}
where the scalar field $\phi$ is derived from $f'({\cal R})$, the first derivative in $\cal R$ of $f(\cal R)$.

In the weak gravitational field limit, it is possible to demonstrate that Newtonian potential for Hybrid Gravity is given as \cite{bork16a}:

\begin{equation}
\Phi(r) = -\dfrac{G}{1 + \phi_0} \left[1 - \dfrac{\phi_0}{3} e^{-m_\phi r} \right] \dfrac{M(r)}{r},
\label{equ23}
\end{equation}

\noindent where the main parameters of Hybrid Gravity are $m_\phi$ and $\phi_0$, derived from the self-interaction potential $V(\phi)$.

For modeling the stellar kinematics in the elliptical galaxies, we assumed that the mass distribution within them may be described by the singular isothermal sphere (SIS) model: $M(r) = 2\sigma_{SIS}^2\,G^{-1}\,r$, like in our previous paper \cite{capo20}.

\section{Hybrid Gravity Fundamental Plane}
\label{sec03}
Hybrid Gravity parameters can be constrained by the FP.

Let us label the Newtonian potential with $\Phi_N(r)$ and circular velocity in this potential with $v_N(r)$. Then, $\Phi_N(r) = -\dfrac{GM(r)}{r}$ and $v_N^{2}(r) = r \cdot \Phi_N^\prime(r)$. In the case of hybrid modified potential, supposing the spherically distributed mass in elliptical galaxies, for circular velocity we have $v_c^{2}(r) = r \cdot \Phi^\prime(r)$. Here we start from the Hybrid Gravitational potential \eqref{equ23} and derive the connection between $v_c^{2}(r)$ and parameters of this potential. So we start from Eq. (\ref{equ23}) and its derivative, and then we obtain:

\begin{eqnarray}
\Phi(r) &=& -\dfrac{GM(r)}{(1 + \phi_0)r} + \dfrac{\phi_0}{3(1 + \phi_0)} \dfrac{GM(r)}{r} e^{-m_\phi r}; \nonumber \\[0.2cm]
\Phi^\prime(r) &=& -\dfrac{1}{1 + \phi_0}\left(\dfrac{GM(r)}{r}\right)^\prime + \nonumber \\
&&-\dfrac{\phi_0}{3(1 + \phi_0)} \left(-\dfrac{GM(r)}{r}\right)^\prime e^{-m_\phi r} + \nonumber \\
&&-\dfrac{\phi_0}{3(1 + \phi_0)} \left(-\dfrac{GM(r)}{r}\right) \left(e^{-m_\phi r}\right)^\prime \nonumber \\[0.2cm]
&=& \dfrac{1}{1 + \phi_0} \Phi_N^\prime(r) - \dfrac{\phi_0}{3(1 + \phi_0)} \Phi_N^\prime(r) e^{-m_\phi r} + \nonumber \\
&&-\dfrac{\phi_0}{3(1 + \phi_0)} \Phi_N(r) (-m_\phi) e^{-m_\phi r}; \nonumber \\[0.2cm]
r\Phi^\prime(r) &=& \dfrac{1}{1 + \phi_0}r \Phi_N^\prime(r) - \dfrac{\phi_0}{3(1 + \phi_0)}r \Phi_N^\prime(r) e^{-m_\phi r} + \nonumber \\
&&+\dfrac{\phi_0 m_\phi}{3(1 + \phi_0)}r \Phi_N(r) e^{-m_\phi r}.
\label{equ31}
\end{eqnarray}

\begin{eqnarray}
v_c^{2}(r) &=& \dfrac{1}{1 + \phi_0} v_N^{2}(r) -\dfrac{\phi_0}{3(1 + \phi_0)} v_N^{2}(r) e^{-m_\phi r} + \nonumber \\
&&-\dfrac{\phi_0 m_\phi r}{3(1 + \phi_0)} v_N^{2}(r) e^{-m_\phi r} \nonumber \\[0.2cm]
&=& \dfrac{v_N^2(r)}{1 + \phi_0} \left[1 - \dfrac{\phi_0}{3} \left(m_\phi r + 1 \right) e^{-m_\phi r} \right].
\label{equ32}
\end{eqnarray}
Having in mind that the Newtonian circular velocity at the effective radius, i.e. for $r = r_e$, is $v_N(r_e) = \sigma_0$, where $\sigma_0$ is the observed velocity dispersion see Table 1 in Ref. \cite{burs97} in relation to our considered sample of elliptical galaxies), the circular velocity at the $r_e$ can be written in the following form:

\begin{equation}
v_c^2(r_e) = \dfrac{\sigma_0^2}{1 + \phi_0} \left[1 - \dfrac{\phi_0}{3}(m_\phi r_e + 1) e^{-m_\phi r_e} \right].
\label{equ33}
\end{equation}

\noindent Furthermore, let us introduce the following variable: $w = m_\phi r_e.$

Then, one can obtain the following $v_c/\sigma$ relation at $r_e$ for the ellipticals in Hybrid Gravity:

\begin{equation}
\dfrac{v_c}{\sigma} = \dfrac{v_c(r_e)}{\sigma_0} = \sqrt{\dfrac{1}{1 + \phi_0} \left[1 - \dfrac{\phi_0}{3}(w + 1) e^{-w}\right]}.
\label{equ34}
\end{equation}

\begin{figure*}[ht!]
\centering
\includegraphics[width=0.80\textwidth]{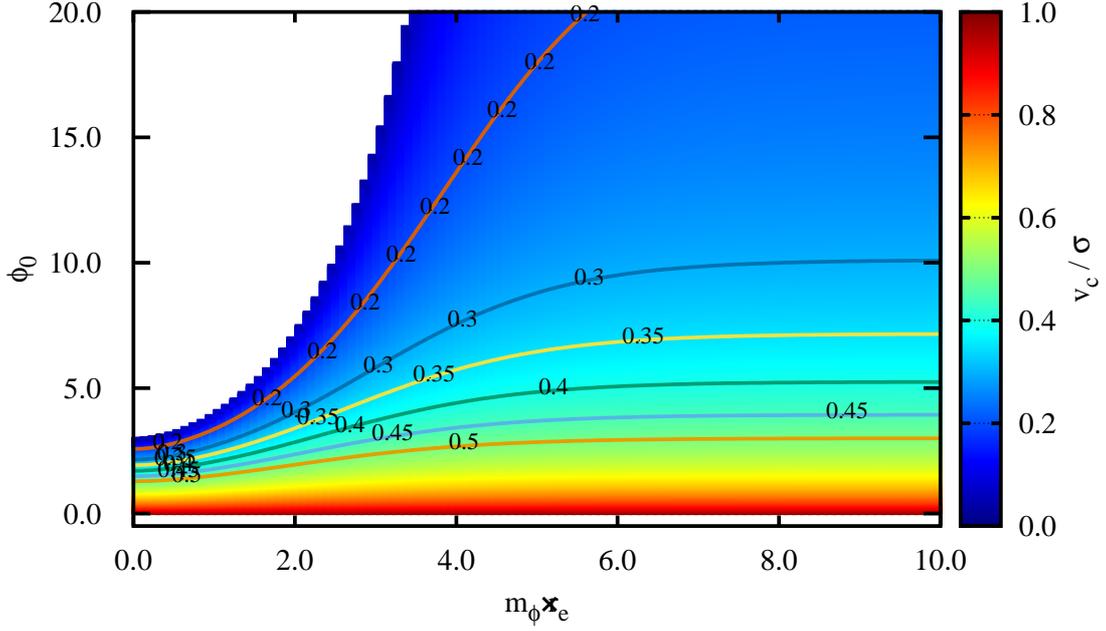}
\caption{The $v_c/ \sigma$ relation for the elliptical galaxies in the Hybrid Gravity represented by different color shades, as well as its dependence on the $(m_\phi \cdot r_e,\phi_0)$ parameter space. Values for $v_c/ \sigma$ are designated by lines for 0.2, 0.3, 0.35, 0.4, 0.45 and 0.5, respectively.}
\label{fig01}
\end{figure*}

\begin{figure*}[ht!]
\centering
\includegraphics[width=0.80\textwidth]{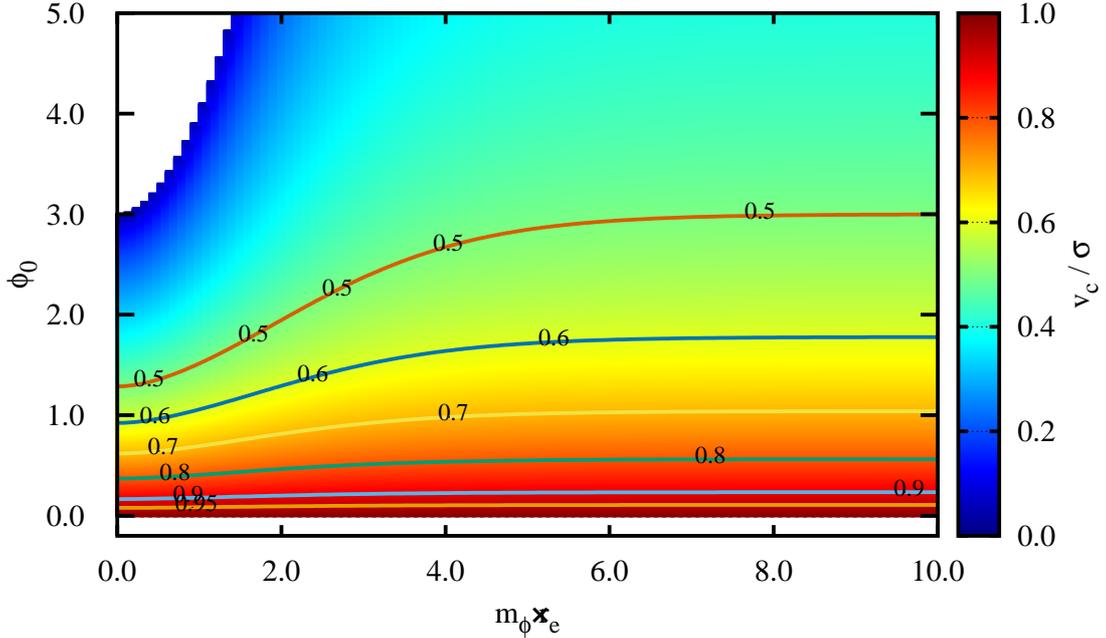}
\caption{The same as in Fig. \ref{fig01}, but for the following values of $v_c/ \sigma$ relation: 0.5, 0.6, 0.7, 0.8, 0.9 and 0.95.}
\label{fig02}
\end{figure*}

\begin{figure*}[ht!]
\centering
\includegraphics[width=0.80\textwidth]{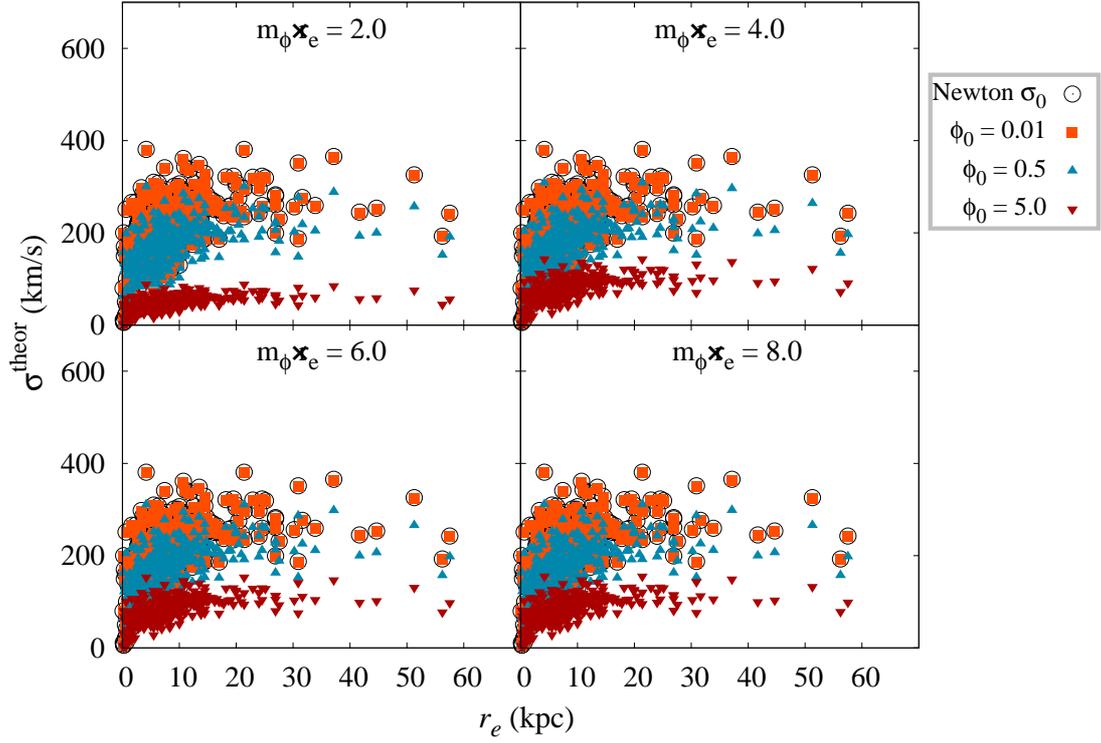}
\caption{Velocity dispersion $\sigma^{theor}$ as a function of the effective radius $r_e$ for elliptical galaxies, for four different values of the $m_\phi \cdot r_e$ product: 2, 4, 6 and 8. The Newtonian velocity dispersion at the effective radius $\sigma_0$ is taken from \cite{burs97}. Theoretical values of velocity dispersion $\sigma^{theor}$ are calculated for the three values of Hybrid Gravity parameter $\phi_0$: 0.01, 0.5 and 5.0.}
\label{fig03}
\end{figure*}

\begin{figure*}[ht!]
\centering
\includegraphics[width=0.80\textwidth]{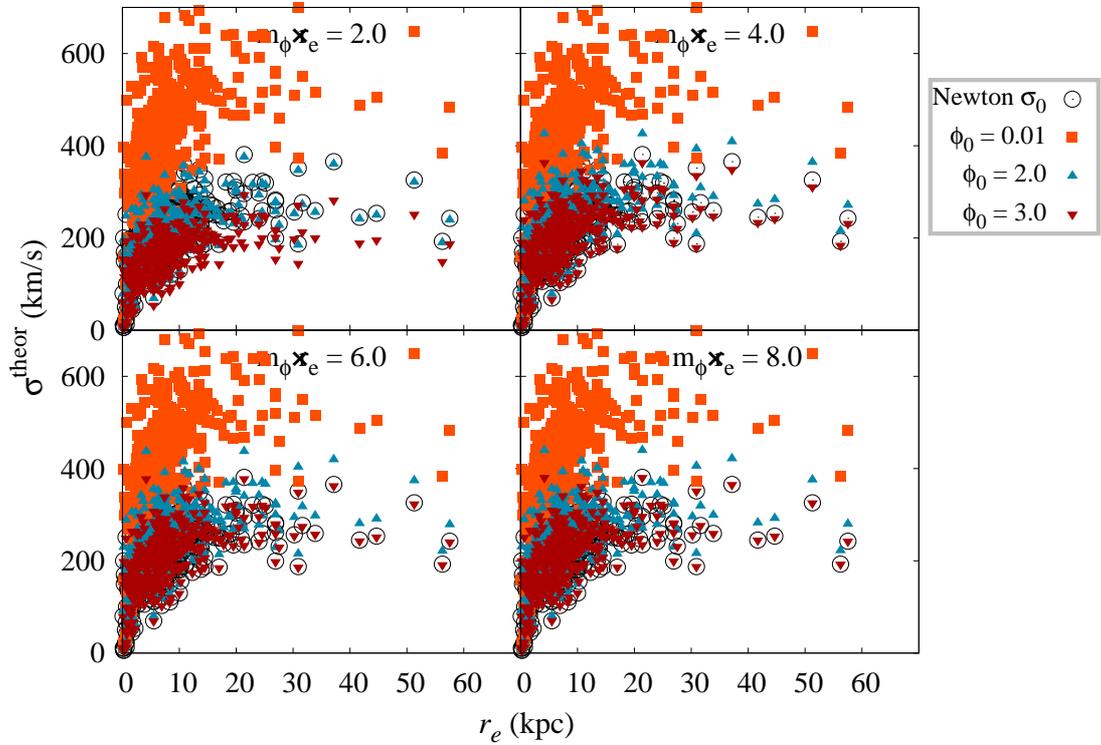}
\caption{The same as in Fig. \ref{fig03}, but for $\sigma^{theor}(r_e) = 0.5\times v_c(r_e)$ and the following three values of $\phi_0$: 0.01, 2.0 and 3.0.}
\label{fig04}
\end{figure*}

\begin{figure*}[ht!]
\centering
\includegraphics[width=0.80\textwidth]{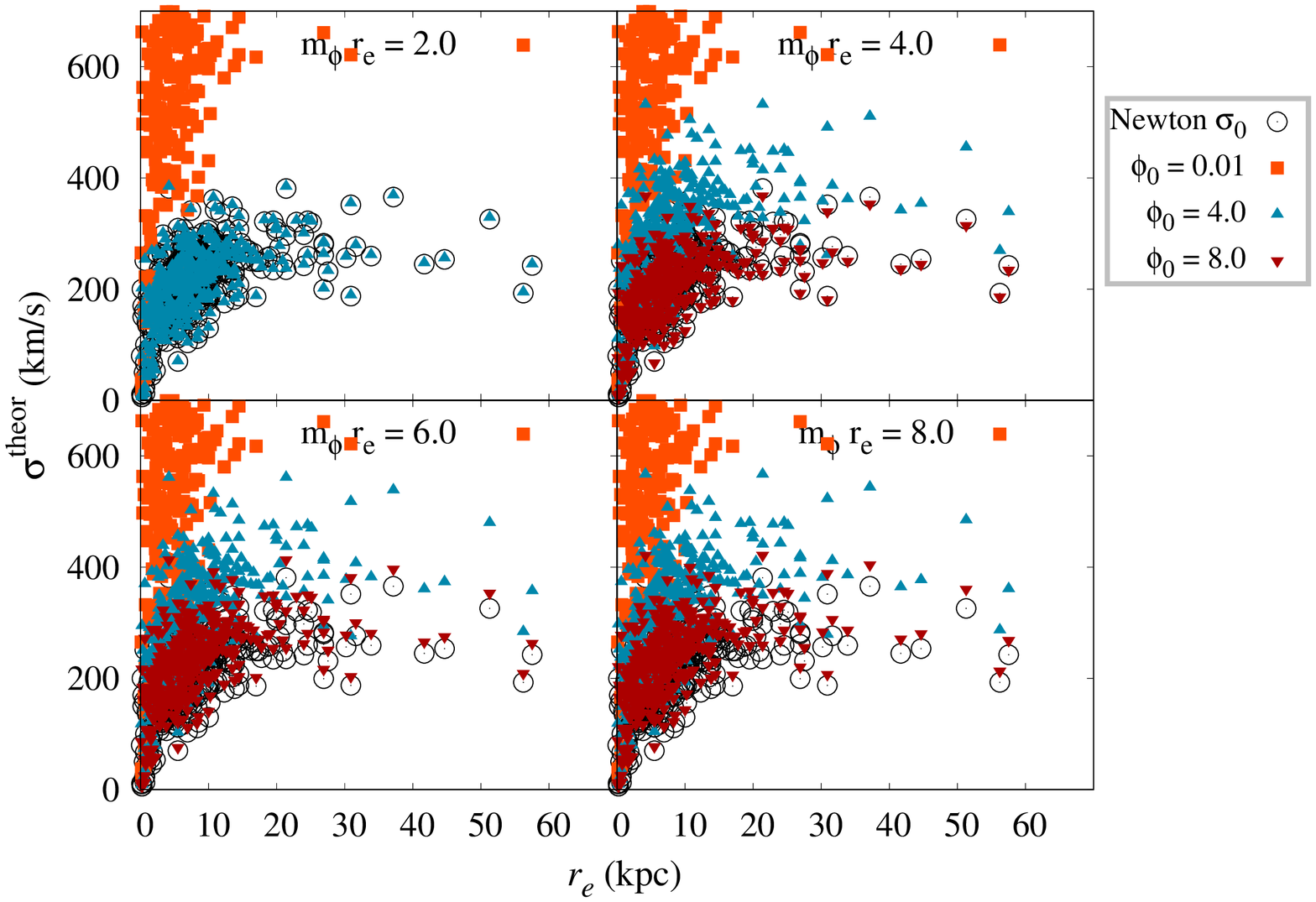}
\caption{The same as in Figs. \ref{fig03} and \ref{fig04}, but for $\sigma^{theor}(r_e) = 0.3\times v_c(r_e)$ and the following three values of $\phi_0$: 0.01, 4.0 and 8.0.}
\label{fig05}
\end{figure*}

\section{Results and discussion}
\label{sec04}

Let us study now the $v_c/ \sigma$ relation in the $(m_\phi,\phi_0)$ parameter space of Hybrid Gravity. We constrain the parameters using a sample of elliptical galaxies given in \cite{burs97}. Also, we study velocity dispersion $\sigma^{theor}$ as a function of the effective radius $r_e$. In order to check how different kinematical properties influence the Hybrid Gravity parameters, we constrain the gravitational parameters with the following values of $v_c/\sigma$ relation: 1, 0.5 and 0.3. One should have in mind that, in the observed sample, $\sigma_0$ is equal to the Newtonian circular velocity. The method that we are using is described in detail in references \cite{bork16a,capo20,bork16b,bork19,bork20} and references therein.

Fig. \ref{fig01} shows $v_c/ \sigma$ relation for elliptical galaxies in Hybrid Gravity representation in the $(m_\phi \cdot r_e,\phi_0)$ parameter space. The values of $v_c/ \sigma$ relation are calculated numerically and represented by different color shades. Some chosen values for $v_c/ \sigma$ are designated by lines: 0.2, 0.3, 0.35, 0.4, 0.45 and 0.5, respectively. For a specific value of the $v_c/ \sigma$ relation, these lines represent the values of the parameters for which we expect good agreement between theoretical predictions and observations. For example, if we choose the value $v_c/ \sigma$ = 0.3, in the area along line designated with 0.3 and in nearby region, the agreement between theory and observations is very good. If the parameters $(m_\phi\cdot r_e,\phi_0)$ are more scattered with respect to the designated lines, then this agreement will be worse. We can see that, if we increase the value of the parameter $m_\phi\cdot r_e$ in region from 0 to 6, the values for parameter $\phi_0$ change drastically. It means that, in this region, for fixed values of parameter $\phi_0$, values of the $v_c/ \sigma$ relation is very sensitive on parameter $m_\phi\cdot r_e$. For values $m_\phi\cdot r_e \gtrsim 6$, the lines become almost horizontal for all studied values of parameter $\phi_0$. It means that, in this region, for a fixed value of parameter $\phi_0$, values of the $v_c/ \sigma$ relation are not influenced by parameter $m_\phi\cdot r_e$. 

Fig. \ref{fig02} shows the $v_c/ \sigma$ relation for elliptical galaxies in the Hybrid Gravity like Fig. \ref{fig01}, but for different values of parameter $\phi_0$ and $v_c/ \sigma$ relation: the values of $v_c/ \sigma$ are designated by lines for 0.5, 0.6, 0.7, 0.8, 0.9 and 0.95, respectively. We can notice the same tendency like in Fig. \ref{fig01}, i.e. that if we change the value of parameter $m_\phi\cdot r_e$, in region from 0 to 3, the corresponding values for parameter $\phi_0$ change drastically. The values of $v_c/ \sigma$ relation are more influenced by increasing of $m_\phi\cdot r_e$ for smaller values of $v_c/ \sigma$. For values $m_\phi\cdot r_e \gtrsim 3$, studied lines become almost horizontal for all six values of $v_c/ \sigma$ ratio. The main difference from Fig. \ref{fig01} is that in this case saturation of $v_c/ \sigma$ relation happens for $m_\phi\cdot r_e \gtrsim 3$, since that the corresponding curve becomes constant. This means that for higher values of $v_c/ \sigma$ ratio its saturation could be expected for smaller values of $m_\phi\cdot r_e$ product.

Fig. \ref{fig03} shows theoretical values for the velocity dispersion of elliptical galaxies $\sigma^{theor}$, which is assumed to be equal to $v_c(r_e)$ as a function of the effective radius $r_e$ for elliptical galaxies. The values for $\sigma^{theor}$ are presented for 4 different products of $m_\phi\cdot r_e$: 2, 4, 6 and 8. In the same figure, it is presented the Newtonian velocity dispersion at the effective radius $\sigma_0$ \cite{burs97}. Theoretical values of velocity dispersion $\sigma^{theor}$ are calculated for the three different values of Hybrid Gravity parameter $\phi_0$: 0.01, 0.5 and 5.0. We can conclude that in all 4 cases for the $m_\phi\cdot r_e$ product, the agreement between theoretical results and astronomical observation is excellent in case of Hybrid Gravity parameter $\phi_0$ = 0.01. In case of $\phi_0$ = 0.05,  agreement is not so good, but in case of $\phi_0$ = 5.0 it is very poor. If we look at Fig. \ref{fig02}, we can notice lines (the $v_c/ \sigma$ ratio) very near to the value of 1 (0.9 and 0.95) and we can conclude that parameter $\phi_0$ should be very small in that case. Also, we can notice that obtained result is almost not sensitive to the value of $m_\phi\cdot r_e$ product.  

Fig. \ref{fig04} represents the same as in Fig. \ref{fig03}, but for $\sigma^{theor}(r_e)$ = $0.5\times v_c(r_e)$ and the following three values of $\phi_0$: 0.01, 2.0 and 3.0. We can conclude that, in all 4 cases, of the $m_\phi\cdot r_e$ product, the agreement between theoretical results and astronomical observation is very poor in case of $\phi_0$ = 0.01. In case of $\phi_0$ = 2.0 and 3.0 agreement is much better. For the value of $m_\phi\cdot r_e$ = 2, agreement is better in case $\phi_0$ = 2.0, and for the values of $m_\phi\cdot r_e$ = 4, 6 and 8 agreement is better in case $\phi_0$ = 3.0. If we look at Fig. \ref{fig02}, we can conclude that this behavior is expected (notice curve that represents the $v_c/ \sigma$ ratio = 0.5 and its position in the $(m_\phi\cdot r_e,\phi_0)$ parameter space). 

Fig. \ref{fig05} represents the same as in Figs. \ref{fig03} and \ref{fig04}, but for $\sigma^{theor}(r_e) = 0.3\times v_c(r_e)$ and the following three values of $\phi_0$: 0.01, 4.0 and 8.0. We can see that, in all 4 cases of the $m_\phi\cdot r_e$ product, the agreement between theoretical results and observations is very poor in case of $\phi_0$ = 0.01. In case of $\phi_0$ = 4.0 and for the value of $m_\phi\cdot r_e$ = 2, agreement is excellent. Also in case $\phi_0$ = 8.0, and for the value of $m_\phi\cdot r_e$ = 8, agreement is satisfactory. For the values of $m_\phi\cdot r_e$ = 4 and 6, the agreement is also satisfactory. If we look position of line designated by 0.3 in the $(m_\phi\cdot r_e,\phi_0)$ parameter space (see Fig. \ref{fig01}), we can conclude that these results are expected.

For all the three studied values of $\sigma^{theor}$, a good agreement between theoretical calculations and astronomical observations is achieved for different values of Hybrid Gravity parameter $\phi_0$. We can conclude that the parameter $\phi_0$ is very sensitive on the value of $v_c/\sigma$ relation.

\section{Conclusions}
\label{sec05}

We studied the $v_c/ \sigma$ relation in the $(m_\phi \cdot r_e,\phi_0)$ parameter space of Hybrid  Gravity. We constrain its parameters using a sample of elliptical galaxies given in \cite{burs97}. 

According to the above results, we can conclude that the Hybrid  Gravity is able to explain elliptical galaxies with different stellar kinematics described by $v_c/\sigma$ relations shown in Figs. \ref{fig01} and \ref{fig02}, without  introducing the dark matter hypothesis. For theoretical values $v_c/\sigma$ = 1, we determine parameters $\phi_0$ and $m_\phi$ of Hybrid Gravity, since the same condition was assumed at the observed sample for the Newtonian gravity \cite{burs97}. In order to get good agreement between numerical calculation and observations, we obtain that the value parameter $\phi_0$ has to be close to $\precsim$0.01 for all values of the product $m_\phi\cdot r_e$ in the studied range. In order to check how different kinematical properties influence the  parameters, we also tested what happens if a given system of elliptical galaxies is kinematically hotter. We decrease the value of $v_c/\sigma$ to 0.5 and to 0.3, and the best agreement was obtained for larger value of parameter $\phi_0$. Also, we can notice that, in these cases (kinematically hotter systems), the influence of $m_\phi\cdot r_e$ product should be taken into account. From Figs. \ref{fig03}-\ref{fig05} we can notice that the Hybrid Gravity parameter $\phi_0$ is very sensitive to the ratio $v_c/\sigma$. Therefore, the $v_c/\sigma$ relation and the FP can be used as standard tools to probe the Hybrid Gravity parameters in the weak gravitational field limit, as well as to constrain these parameters. Clearly, the same procedure can be adopted to test other theories of gravity.

\paragraph{Acknowledgments}
This work is supported by Ministry of Education, Science and Technological Development of the Republic of Serbia. P.J. wishes to acknowledge the support by this Ministry through the Project contract No. 451-03-9/2021-14/200002. SC acknowledges the support of {\it Istituto Nazionale di Fisica Nucleare} (INFN), {\it iniziative specifiche} MOONLIGHT2 and QGSKY.

\paragraph{Authors contributions}
All coauthors participated in calculation and discussion of obtained results. The authors contributed equally to this work.

\paragraph{Data Availability Statement}
This manuscript has no associated data or the data will not be deposited. [Authors' comment: All relevant data are in the paper.]

\end{document}